\documentclass[a4paper,11pt]{article}

\usepackage{pos}
\usepackage{lipsum} 
\usepackage{wrapfig}
\usepackage{multirow}

\title{An updated list of target sources for IceCube neutrino cluster alerts}

\ShortTitle{Updated source list for IceCube cluster-alerts}

\author{The IceCube Collaboration \\{\normalsize \normalfont(a complete list of authors can be found at the end of the proceedings)}\\}

\emailAdd{caterina.boscolomeneguolo@studenti.unipd.it}
\emailAdd{elisa.bernardini@unipd.it}
\emailAdd{jeanpierre.jonckheere@studenti.unipd.it}
\emailAdd{sarah.mancina@icecube.wisc.edu}

\abstract{

Multimessenger astronomy seeks to uncover the origins of cosmic rays and neutrinos. The IceCube Neutrino Observatory plays a key role in monitoring the sky for revealing high energy neutrinos and neutrino time clusters possibly associated with astrophysical sources, issuing alerts to the astrophysical community for significant excesses. This enables joint observations with other astronomical facilities that could reveal the hidden mechanisms behind the most extreme environments in the Universe. In particular, since 2006 the Gamma-ray Follow-Up (GFU) program shares cluster alerts with partner Imaging Air Cherenkov Telescopes. 

The faint cosmic signals, searched against large atmospheric backgrounds, are widely masked by the statistical penalties that arise when scanning the full sky in an unbiased way. Hence, targeted analyses of pre-selected neutrino source candidates have proven to increase our search sensitivity.

Our understanding of astrophysical environments has improved in recent years, with evidence of neutrino emission from the blazar TXS~0506+056 and the Seyfert galaxy NGC~1068. The aim of expanding observational possibilities and engaging the broader scientific community through public cluster alerts has motivated the creation of a new list of target sources to be monitored by IceCube. This contribution presents the systematic compilation of this list, which extends the well-established focus on gamma-ray bright active galactic nuclei (AGN) to include X-ray bright AGN and binary systems.

\vspace{4mm}

{\bfseries Corresponding authors:}
Caterina Boscolo Meneguolo$^{1,2*}$, 
Elisa Bernardini$^{1,2}$, 
Jean-Pierre Jonckheere$^{1}$,
Sarah Mancina$^{1,2}$\\
{$^{1}$ \itshape Dipartimento di Fisica e Astronomia Galileo Galilei, Università Degli Studi di Padova}\\
{$^{2}$ \itshape INFN Padova}\\[4mm]
$^*$ Presenter
}

\ConferenceLogo{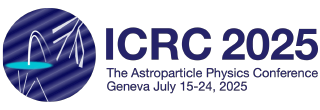}

\FullConference{39th International Cosmic Ray Conference (ICRC2025)\\
 15–24 July 2025\\
Geneva, Switzerland\\}

\begin{document}

\maketitle
\section{Introduction}\label{sec:intro}
Since the discovery of a diffuse flux of astrophysical neutrinos~\cite{IceCube:2013low}, IceCube has been actively investigating their origins. 
As neutrinos are produced in hadronic interactions and can travel straight and unabsorbed across the universe, they are powerful probes of cosmic ray (CR) accelerators~\cite{AHLERS201873}. 

Recent results have advanced our understanding of such accelerators. The evidence of neutrino emission from the Seyfert galaxy NGC~1068~\cite{IceCube:2022NGC} has strenghtened the interest toward active galactic nuclei (AGN) as promising candidates, while an excess of neutrinos from the Galactic Plane~\cite{IC2023gal_plane} points to potential sources within our Galaxy. Despite these observations, the origin of the diffuse flux is still largely unknown, since these sources can only account for a portion of it.

Multimessenger astrophysics aims to uncover the nature of extreme cosmic environments through the combined observations of different cosmic messengers, such as neutrinos and photons. 
However, detecting high-energy gamma rays associated with neutrino emission is challenging due to absorption by extragalactic background light (EBL), as well as the limited duty cycle and field of view of Imaging Air Cherenkov Telescopes (IACTs). These instruments benefit from real-time alerts that trigger follow-up observations of candidate transient phenomena.
IceCube plays a crucial role in this context, thanks to its 99\% uptime, full-sky coverage and a robust real-time infrastructure that sends alerts to the astrophysical community with minimal delay (typically 30-40~s)~\cite{Aartsen_2017online, Abbasi_2023IceCat}. 
This approach enabled in 2017 the landmark joint detection of a high-energy neutrino and enhanced gamma-ray emission from the blazar TXS~0506+056~\cite{TXS_2017}, identifying the first candidate for an astrophysical source of high-energy ($\sim$TeV) neutrinos with a $\sim 3 \sigma$ significance.

The Gamma-ray Follow-Up (GFU) program~\cite{Kintscher2020Rapid} has contributed to the multi-messenger effort starting in 2006 at the precursor detector AMANDA-II, and later evolving and being incorporated into IceCube. It issues alerts based on two main strategies.  The first selects neutrino candidates with exceptionally high energy, standing out from the softer atmospheric spectrum. The second identifies spatial and temporal clusters of neutrino events, which may signal a flaring source emerging from the uniform background. Both approaches help in the identification of cosmic neutrinos in IceCube, which is particularly challenging due to the dominant atmospheric background. This work focuses on the GFU-cluster algorithm.

The GFU-cluster search operates in two modes: first, an unbiased scan of the entire sky searches for clustering of neutrino events by dividing the sky into pixels. However, the repeated test of a large number of pixels introduces a large trial factor, which limit the sensitivity. To address this issue, a second approach of the GFU-cluster search also monitors a pre-selected list of targeted sources, significantly improving the discovery potential.

Since 2019, GFU-cluster alerts have been shared with partner IACTs under a memorandum of understanding. The improved insights into neutrino sources and the will to engage open collaboration motivate a major upgrade: transitioning GFU-cluster alerts to public distribution via NASA General Coordinates Network (GCN)\footnote{\href{https://gcn.nasa.gov/}{https://gcn.nasa.gov/}} notices. 
Details are provided in Ref.~\cite{Mancina:2025icrc}. This, in turn, calls for an updated list of monitored sources. A description of GFU-cluster alerts is given in Sec.~\ref{sec:GFU} while the new selection of targeted sources is presented in Sec.~\ref{sec:newlist}. A summary of the final selection is given in Sec.~\ref{sec:results} and conclusions are presented in Sec.~\ref{sec:conclusion}.

\section{GFU-cluster alerts}\label{sec:GFU}
GFU-cluster alerts, in their current form, have been shared privately under a memorandum of understanding with four IACTs: H.E.S.S.~\cite{aharonian2006observations}, MAGIC~\cite{2012APh....35..435A}, and VERITAS~\cite{Holder:2006gi} since 2019 and LST-1~\cite{Abe_2023} since 2023.
Details on the GFU-cluster alert system are presented in Ref.~\cite{Boscolo:2023icrc}. Here, we will only provide a brief overview of the algorithm followed by a description of the current list of monitored sources.

\subsection{The algorithm}\label{sec:algorithm}
\begin{wrapfigure}{r}{0.55\textwidth}
\centering
\vspace{-40pt}
\includegraphics[width=\linewidth]{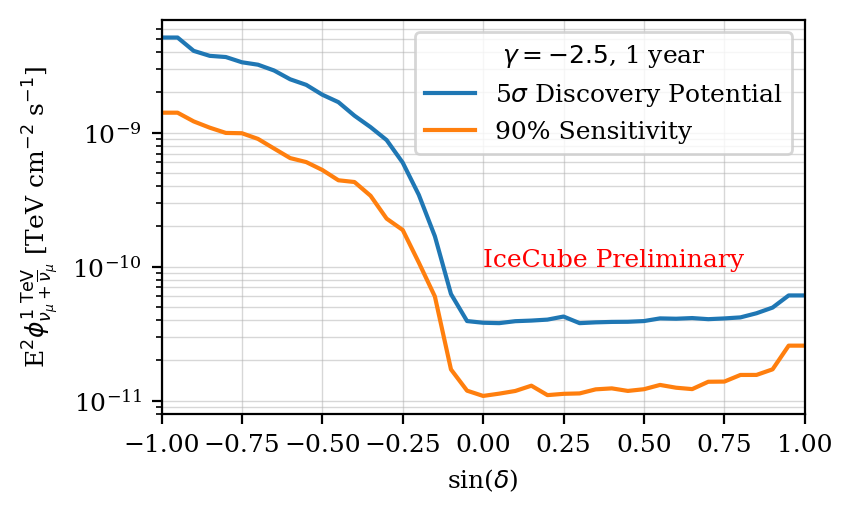}
\caption{Sensitivity and discovery potential for time-integrated point-source analysis for one year, assuming a power-law spectrum of the form $dN/dE\propto E^{\gamma}$ with spectral index $\gamma=-2.5$.}\label{fig:sens}
\vspace{-10pt}
\end{wrapfigure}
The GFU-cluster system performs a likelihood-based analysis of muon neutrino events selected by an optimized online reconstruction and selection pipeline~\cite{Kintscher2020Rapid}. Each event is evaluated against a set of source hypotheses, either sky pixels (in all-sky mode) or a predefined list of monitored source coordinates. This work focuses exclusively on the source-list mode.

For each tested source, a signal-to-background (S/B) weight is calculated for the incoming event, based on its energy and closeness to the source. If any source yields S/B > 1, the likelihood analysis is triggered. The algorithm then scans several time windows, looking back in time up to 180 days, to identify clusters of spatially correlated events. The window yielding the highest likelihood is selected. If the resulting cluster candidate exceeds a predefined significance threshold, an alert is issued. 

The alert threshold for the source-list mode is set at 3$\sigma$ (3.5$\sigma$ for VERITAS), corresponding to a false alert rate of $\sim$10 alerts per telescope per year. 
The sensitivity and discovery potential of the time-integrated point-source analysis~\cite{Aartsen_2017_7yearPS} for one year of data is shown in Fig.~\ref{fig:sens}. A power-law spectrum of the form $dN/dE\propto E^{\gamma}$ with a spectral index $\gamma=-2.5$ is assumed for the neutrino flux, matching recent results on the observed astrophysical spectrum~\cite{globalfit:2023icrc}. Details on the sensitivity and discovery potential calculation can be found in Refs.~\cite{Aartsen_2017_7yearPS} and~\cite{Kintscher2020Rapid}.

\subsection{The list of monitored sources}\label{sec:list_today}
The current GFU-cluster alert stream relies on three independent source lists, compiled between 2017 and 2019 in agreement with the partner IACTs. Each source in every list is evaluated as an independent hypothesis. If a cluster surpasses the alert threshold, a notification is sent to the IACT associated with the triggering source list.
These lists were drawn from selections of gamma-ray bright extragalactic sources in the Third Fermi-LAT catalog of AGN (3LAC)~\cite{Ackermann_2015_3lac} and the Third Catalog of Hard Fermi-LAT Sources (3FHL)~\cite{Ajello_2017_3fhl}. The selection criteria were designed to favor variable candidate neutrino emitters and maximize the likelihood of detection by IACTs.

Sources were required to exhibit >99\% C.L. variability~\cite{Ackermann_2015_3lac}, have a measured redshift below 1, and reach a sufficient elevation above the telescope horizon (>45°–60°). Each sources' flux was extrapolated above 100 GeV and corrected for EBL absorption, and then multiplied by a factor of 10 to simulate a potential flare. The resulting extrapolated flux was then verified to be within the telescope’s 5$\sigma$ discovery potential for a fixed observation time (2 or 5 hours). Additionally, each IACT manually included a few selected targets (e.g., known TeV emitters, the Crab Nebula, the Galactic Center).

VERITAS' list was later updated to include known TeV extragalactic sources and Galactic binaries. The variability cut was removed, and sources were ranked by extrapolated flux weighted by telescope sensitivity at peak elevation. For sources lacking redshift, a default value of $z = 0.3$ was assumed. The sources with the highest ranking were selected for inclusion.

The resulting lists contain 179 sources for MAGIC (also adopted by LST-1), 190 for VERITAS, and 139 for H.E.S.S., with substantial overlap. In total, 339 unique sources are included, of which 129 appear in more than one list. Coordinates were selected per IACT preference, resulting in minor positional discrepancies for 97 shared sources, with a mean difference of 0.02°, enough to affect cluster significance and alert issuance in some cases. Source positions are shown as gray dots in Fig.~\ref{fig:new_list}, together with coordinates of the sources in the updated list, described in Sec.~\ref{sec:newlist}, which are the main focus of this work.

Additional details and ongoing upgrades to the system are described in Ref.~\cite{Mancina:2025icrc}.

\section{An updated source list for GFU-cluster alerts}\label{sec:newlist}
An  update of the monitored source catalog is motivated by improved insights into potential neutrino sources,  along with the transition toward public GFU-cluster alerts aimed at engaging a broader community of follow-up observatories (e.g., X-ray telescopes).
To this end, a unified source list was compiled, removing IACT-specific observability constraints in favor of criteria focused on IceCube sensitivity and scientific relevance, to be fully maintained by IceCube.

Sources were grouped into two categories \textit{a priori}, based on the expected energy range of their associated gamma-ray emission: a \textit{GeV gamma-ray motivated selection} (Sec.~\ref{sec:GeV}) and a \textit{keV X-ray motivated selection} (Sec.~\ref{sec:keV}).

For each category, relevant source catalogs were used as input, and sources were selected based on a figure of merit (FoM): the ratio of their electromagnetic flux in the catalog energy band as a proxy of neutrino emission, to the sensitivity at the source position (as introduced in Sec.~\ref{sec:algorithm} and Fig.~\ref{fig:sens}). This metric enables internal ranking, acknowledging that overlaps across the selections are natural, as many candidate neutrino sources emit across multiple energy bands.

The resulting list will be complemented by a \textit{TeV gamma-ray motivated selection}, which is under preliminary construction at the time of writing. This sublist will include both Galactic and extragalactic sources detected in the TeV range, or expected to emit in the TeV range based on their extrapolated GeV flux, and will maintain a focus on IACT detectability in a similar way to the old source selection described in Sec.~\ref{sec:list_today}. Indeed, neutrinos and very high-energy (VHE, $E > 100 \text{~GeV}$) gamma rays are both expected products of CR acceleration via $pp$ or $p\gamma$ interactions. Detecting VHE gamma rays coincident with a neutrino cluster would serve as a clear signature of such interactions. 

The motivation and criteria for selecting candidate neutrino sources that are bright at lower energy bands are explained in the following.

\subsection{GeV gamma-ray motivated selection}\label{sec:GeV}
The contribution of VHE sources to the measured neutrino flux must be limited, to ensure that the associated gamma-ray emission does not overshoot the diffuse gamma-ray background, especially for extragalactic sources. However, AGN, particularly blazars, remain strong candidates supported by the case of TXS~0506+056~\cite{TXS_2017} and their well-characterized, variable GeV emission observed by Fermi-LAT.

The third release of the Fourth Fermi-LAT catalog of AGN (4LAC-DR3) ~\cite{Ajello_2022_4lacdr3} is taken as input for the selection of 200 GeV-bright AGN. A FoM is calculated from the source integral photon flux $(1-100~\text{GeV})$, and the top 200 sources are selected. The photon flux is preferred over the energy flux in order to avoid biasing the ranking toward sources with a hard spectrum, which are preferred in the TeV-based selection under construction. No cut was set on the redshift to include promising neutrino emitters even with large or unmeasured redshift. No specific requirement was set for source variablity, but the selection increases the sources variable at a 99\% C.L.~\cite{Abdollahi_2020_4FGL}  to a portion of $\sim95\%$. 

\subsection{keV X-ray motivated selection}\label{sec:keV}
The neutrino detection from NGC~1068~\cite{IceCube:2022NGC} suggests an optically thick production environment where GeV gamma rays are reprocessed to the keV-MeV energy range. 
This motivates the inclusion of hard X-ray bright extragalactic sources, e.g., Seyfert galaxies and X-ray AGN~\cite{Abbasi_2025_xagn, abbasi2024_seyf}. Similarly, evidence of neutrino emission from the Galactic Plane and models of CR acceleration make X-ray-bright Galactic sources, like microquasars~\cite{Abbasi_2022_xbin}, interesting candidates.

A selection of X-ray bright AGN is performed based on the \textit{Swift/BAT} Spectroscopic Survey (BASS) II~\cite{BASS_datarel2} which provides an estimate of the AGN intrinsic X-ray flux ($14-195~\text{keV}$) upon knowing their level of obscuration. We assume the neutrino flux to be proportional to the intrinsic flux and we calculate the FoM based on it. The highest ranked sources are then selected in such a way that, besides a few overlaps with the GeV selected AGN, further 40 are included for monitoring. 

The selction of hard X-ray galactic sources is based instead on the 157-Month \textit{Swift/BAT} hard X-ray catalog\footnote{\href{https://swift.gsfc.nasa.gov/results/bs157mon/}{https://swift.gsfc.nasa.gov/results/bs157mon/}}~\cite{Oh_2018_BATcatalog}, with the FoM calculated from the source observed flux ($14-195~\text{keV}$), as no intrinsic flux is provided in this case. The top 10 sources are then selected, resulting in nine binaries and the Crab nebula.

Finally, a model-driven selection of microquasars is performed, based on the sources listed in the analysis of Distefano et al.~\cite{Distefano_2002} based on an internal shock model~\cite{Levinson2001_model}, 
and the analysis by Kantzas et al.,~\cite{Kantzas2023_paper} based on a multizone jet model~\cite{Kantzas2020_model}. 
The 10 northernmost microquasars are selected to populate the region of the sky where the analysis sensitivity is at its maximum, i.e., $\sin{(\text{dec})}\gtrsim -0.25$. Two additional microquasars are finally added to the selection, in order to include in the list all the microquasars observed at up to $\sim100$ TeV by the Large High Altitude Air Shower Observatory (LHAASO), which are candidate efficient CR accelerators according to Ref.~\cite{LHAASO:2024psv}.

\begin{table}[]
\centering
\begin{tabular}{llllr}
\hline
\multicolumn{1}{l}{\textbf{Selection}} & \textbf{Sub-selection }& \textbf{Catalog} &\textbf{Flux for FoM} & \textbf{Sources} \\ \hline

\multicolumn{1}{c}{\multirow{2}{*}{GeV $\gamma$-rays}} & \multirow{2}{*}{{AGN}} & \multirow{2}{*}{4FGL}
& \multirow{2}{*}{\begin{tabular}[c]{@{}l@{}}Photon flux (1-100 GeV)\end{tabular}} & \multirow{2}{*}{200} \\
\multicolumn{1}{l}{} &  &  &  &  \\ \hline
\multicolumn{1}{c}{\multirow{3}{*}{keV X-rays}} & {AGN} & BASS 
& \begin{tabular}[c]{@{}l@{}}Intrinsic flux (14-195 keV)\end{tabular} & 44 \\ 
\multicolumn{1}{l}{} & {Galactic sources} & Swift/BAT 
& \begin{tabular}[c]{@{}l@{}}Observed flux (14-195 keV)\end{tabular} & 10 \\ 
\multicolumn{1}{l}{} & {Microquasars} &
\cite{Distefano_2002, Kantzas2023_paper, LHAASO:2024psv} 
& -& 12 \\ \hline
&  & & \multicolumn{1}{r}{\textbf{Unique sources}} & \textbf{259} \\ 
\end{tabular}
\caption{Summary of the source selections. The GeV $\gamma$-ray and keV X-ray motivated selections are summarized, along with their respective sub-selections with the input catalogs, the specific flux considered for the FoM calculation, and the number of sources selected within the subselection. No flux was considered for the Microquasar selection, as explained in Sec.~\ref{sec:keV}. The final total count of single sources is also given, subtracting repetitions.}\label{tab:counts}
\end{table}

\section{Final selected source catalog}\label{sec:results}
\begin{figure}[t!]
\centering
\includegraphics[width=0.8\linewidth]{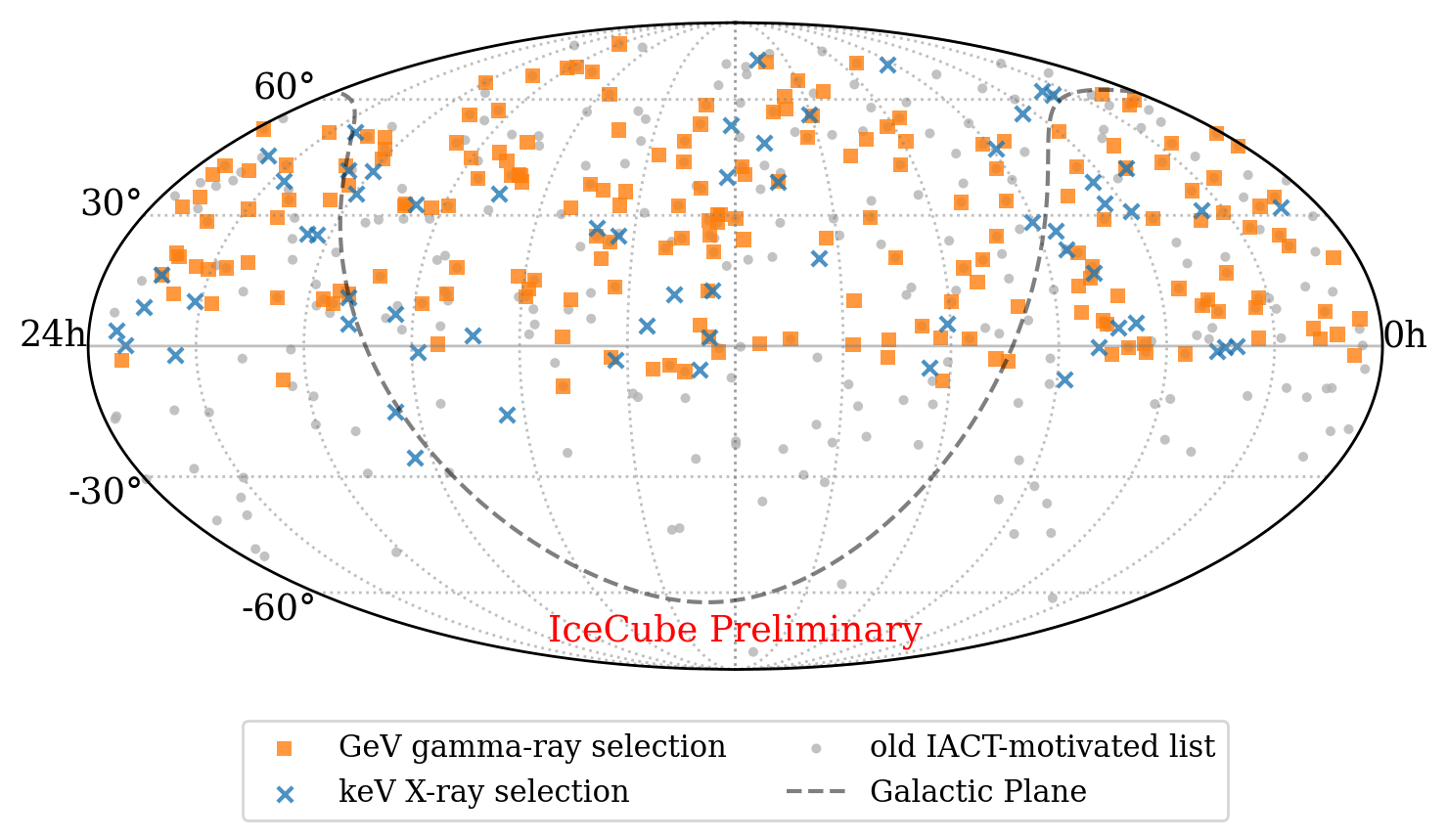}
\caption{Unified list of sources for the GFU-cluster source-list alerts. The sky map is in equatorial coordinates. Sources are marked based on their selection criteria. The old IACT-motivated catalog is shown in gray markers for comparison. The dashed line indicates the location of the Galactic Plane}\label{fig:new_list}
\end{figure}

The selection procedures described in Sec.~\ref{sec:newlist} yield 200 sources motivated by GeV gamma rays and 61 by keV X rays. As expected, there is an overlap between the two selections and the final number of unique sources is 259, which is lower than the current because it leaves room for adding the TeV-motivated sources. A summary of the selection criteria and resulting sources is given in Table~\ref{tab:counts} and a skymap of the unified list is shown in Fig.~\ref{fig:new_list}

Four AGN are in common between the GeV- and keV-motivated selections. These particularly interesting sources are the Seyfert galaxy NGC~1275, the BL Lac blazar Markarian~421 and the flat spectrum radio quasars 3C~454.3 and 3C 273. 
40 Seyfert galaxies are included in the list, which is a novel addition with respect to the previous focus of the cluster alert search on gamma-ray bright blazars. Among them, NGC~1068 is now selected as a promising X-ray bright emitter.

The new list includes 19 Galactic sources, particularly binary systems and microquasars, and the Crab Nebula. 

In view of triggering follow-up observations from a wide range of observatories, the selections that motivated the inclusion of each source are communicated along with the alert GCN notices, giving a criterion for raising interest towards the follow-up.

The source coordinates included in the lists are chosen as the most accurate coordinates provided by the 4FGL catalog for the associated counterparts, while for sources without a counterpart in 4FGL, the coordinates were taken from the associated counterparts in the \textit{Swift/BAT} catalog, or from the SIMBAD Astronomical Database\footnote{\href{http://simbad.cds.unistra.fr/simbad/}{http://simbad.cds.unistra.fr/simbad/}}.

Only 34\% of the sources that are currently part of the GFU program are included in the new selection. 
This significant difference is due to the new focus on IceCube sensitivity (see Fig.~\ref{fig:sens}), instead of IACT detectability criteria. IceCube sensitivity, in fact, favors the horizon region and the Northern sky, as can be visualized in the sky map in Fig.~\ref{fig:new_list} by comparing the new selections (color) and the old list (gray). The inclusion of distant blazars (Sec.~\ref{sec:GeV}), X-ray bright AGN opaque in gamma rays and a fraction of galactic sources (Sec.~\ref{sec:keV}) also presents a departure from the old, IACT-tailored list. 

The \textit{TeV gamma-ray motivated selection} in preparation is expected to partially overlap with the selections presented in this contribution and to further select some blazars included in the current GFU list. In fact, like the previous selection criteria, it will retain a focus on IACT detectability.
The final number of individual sources resulting from the combination of the two selections is expected to be around 350, i.e., similar to the present (see Sect.~\ref{sec:list_today}).

\section{Conclusion}\label{sec:conclusion}
The GFU-cluster alert stream has been operating since 2019, serving as a key tool in coordinating follow-up observations with IACTs. The program is now undergoing a shift to publicly sharing alerts for a list of monitored sources, with the goal of engaging a broader network of observatories spanning a wider range of the electromagnetic spectrum.
This transition is motivated not only by the desire for broader multi-messenger collaboration, but also by recent advances in our understanding of potential neutrino sources, including the evidence for neutrino emission from the Seyfert galaxy NGC~1068 and from the Galactic Plane.

In light of these developments, a new unified list of target sources will be introduced. This list combines three selection strategies. The selection prioritizes sources with high electromagnetic output, considered as a proxy of neutrino emission, while accounting for the sensitivity profile of the GFU-cluster search, which is biased toward the Northern sky under a power-law spectrum assumption with a spectral index of $\gamma=-2.5$. The selections motivated by GeV gamma-ray and keV X-ray emission have been presented, yielding a total of 259 sources. This list will be complemented by a third selection motivated by sources' TeV gamma-ray emission, which is under construction at the time of writing. The resulting list is expected to count $\sim$350 sources.

Compared to the previous source catalog, the updated list represents a significant evolution, with only about 35\% overlap between the previous source catalog and the selection presented in this contribution. This change reflects a deliberate broadening of the program's strategy aimed at maximizing the scientific reach of neutrino follow-up efforts. The revised list now incorporates a more diverse set of targets, including Seyfert galaxies, Galactic sources, and newly selected blazars, representing a refined focus for the ongoing monitoring program.

\vspace{-3mm}
\begingroup
\footnotesize
\setlength{\bibsep}{1pt}
\bibliographystyle{ICRC}
\bibliography{references}
\endgroup

\clearpage

\section*{Full Author List: IceCube Collaboration}

\scriptsize
\noindent
R. Abbasi$^{16}$,
M. Ackermann$^{63}$,
J. Adams$^{17}$,
S. K. Agarwalla$^{39,\: {\rm a}}$,
J. A. Aguilar$^{10}$,
M. Ahlers$^{21}$,
J.M. Alameddine$^{22}$,
S. Ali$^{35}$,
N. M. Amin$^{43}$,
K. Andeen$^{41}$
% ...
\\
\\
% ...
$^{10}$ Universit{\'e} Libre de Bruxelles, Science Faculty CP230, B-1050 Brussels, Belgium \\
% ...
$^{16}$ Department of Physics, Loyola University Chicago, Chicago, IL 60660, USA \\
$^{17}$ Dept. of Physics and Astronomy, University of Canterbury, Private Bag 4800, Christchurch, New Zealand \\
% ...
$^{21}$ Niels Bohr Institute, University of Copenhagen, DK-2100 Copenhagen, Denmark \\
$^{22}$ Dept. of Physics, TU Dortmund University, D-44221 Dortmund, Germany \\
% ...
$^{35}$ Dept. of Physics and Astronomy, University of Kansas, Lawrence, KS 66045, USA \\
% ...
$^{39}$ Dept. of Physics and Wisconsin IceCube Particle Astrophysics Center, University of Wisconsin{\textemdash}Madison, Madison, WI 53706, USA \\
% ...
$^{41}$ Department of Physics, Marquette University, Milwaukee, WI 53201, USA \\
% ...
$^{43}$ Bartol Research Institute and Dept. of Physics and Astronomy, University of Delaware, Newark, DE 19716, USA \\
% ...
$^{63}$ Deutsches Elektronen-Synchrotron DESY, Platanenallee 6, D-15738 Zeuthen, Germany \\
$^{\rm a}$ also at Institute of Physics, Sachivalaya Marg, Sainik School Post, Bhubaneswar 751005, India \\
% ...

\subsection*{Acknowledgments}

\noindent
The authors gratefully acknowledge the support from the following agencies and institutions:
USA {\textendash} U.S. National Science Foundation-Office of Polar Programs,
U.S. National Science Foundation-Physics Division,
U.S. National Science Foundation-EPSCoR,
U.S. National Science Foundation-Office of Advanced Cyberinfrastructure,
Wisconsin Alumni Research Foundation,
Center for High Throughput Computing (CHTC) at the University of Wisconsin{\textendash}Madison,
Open Science Grid (OSG),
Partnership to Advance Throughput Computing (PATh),
Advanced Cyberinfrastructure Coordination Ecosystem: Services {\&} Support (ACCESS),
Frontera and Ranch computing project at the Texas Advanced Computing Center,
U.S. Department of Energy-National Energy Research Scientific Computing Center,
Particle astrophysics research computing center at the University of Maryland,
Institute for Cyber-Enabled Research at Michigan State University,
Astroparticle physics computational facility at Marquette University,
NVIDIA Corporation,
and Google Cloud Platform;
Belgium {\textendash} Funds for Scientific Research (FRS-FNRS and FWO),
FWO Odysseus and Big Science programmes,
and Belgian Federal Science Policy Office (Belspo);
Germany {\textendash} Bundesministerium f{\"u}r Forschung, Technologie und Raumfahrt (BMFTR),
Deutsche Forschungsgemeinschaft (DFG),
Helmholtz Alliance for Astroparticle Physics (HAP),
Initiative and Networking Fund of the Helmholtz Association,
Deutsches Elektronen Synchrotron (DESY),
and High Performance Computing cluster of the RWTH Aachen;
Sweden {\textendash} Swedish Research Council,
Swedish Polar Research Secretariat,
Swedish National Infrastructure for Computing (SNIC),
and Knut and Alice Wallenberg Foundation;
European Union {\textendash} EGI Advanced Computing for research;
Australia {\textendash} Australian Research Council;
Canada {\textendash} Natural Sciences and Engineering Research Council of Canada,
Calcul Qu{\'e}bec, Compute Ontario, Canada Foundation for Innovation, WestGrid, and Digital Research Alliance of Canada;
Denmark {\textendash} Villum Fonden, Carlsberg Foundation, and European Commission;
New Zealand {\textendash} Marsden Fund;
Japan {\textendash} Japan Society for Promotion of Science (JSPS)
and Institute for Global Prominent Research (IGPR) of Chiba University;
Korea {\textendash} National Research Foundation of Korea (NRF);
Switzerland {\textendash} Swiss National Science Foundation (SNSF).
\end{document}